\documentclass[preprint,12pt, a4paper]{elsarticle}

\usepackage{amssymb}
\usepackage{hyperref}
\usepackage{booktabs}
\usepackage{amsmath}
\usepackage{algorithm}
\usepackage{algpseudocode}
\usepackage{varwidth}
\usepackage{comment}
\usepackage{etoolbox}
\usepackage{booktabs}
\usepackage{array}
\usepackage{graphicx}
\usepackage{rotating}
\usepackage{makecell}
\usepackage{threeparttable}
\newcolumntype{L}[1]{>{\raggedright\arraybackslash}p{#1}}

\usepackage{pdflscape}

\newcolumntype{L}[1]{>{\raggedright\arraybackslash}p{#1}}

\newcolumntype{Y}{>{\raggedright\arraybackslash}X}

\newcommand{\LongState}[1]{
  \State \parbox[t]{\dimexpr\linewidth-\ALG@thistlm}{\strut #1\strut}
}
\journal{SoftwareX}

\begin{document}
\renewcommand{\labelenumii}{\arabic{enumi}.\arabic{enumii}}

\begin{frontmatter}

\title{DCEDesignSA: A MATLAB-based Graphical User Interface for Discrete Choice Experiment Design Using Simulated Annealing}

\author[maastricht]{Yinfu Liu}
\author[maastricht,calgary]{Yicheng Mao\corref{cor1}}

\address[maastricht]{Data Analytics and Digitalisation, School of Business and Economics, Maastricht University, Tongersestraat 53, 6211 LM Maastricht, The Netherlands}

\address[calgary]{Department of Mathematics and Statistics, University of Calgary, University Drive NW, Calgary, T2N 1N4, Canada}

\cortext[cor1]{Corresponding author.}
\ead{yicheng.mao1@ucalgary.ca}

\begin{abstract}
DCEDesignSA is a freely available MATLAB package for generating Bayesian $\mathcal{D}$-optimal discrete choice experiment designs. It employs Simulated Annealing to efficiently search the design space and maximise the Bayesian $\mathcal{D}$-optimality criterion under user-specified prior distributions. The toolbox features an interactive graphical user interface, enabling researchers without programming expertise to define experimental settings, generate optimal designs, and export survey-ready designs directly to Qualtrics. DCEDesignSA supports interaction terms in utility, no-choice alternatives, and presentation order effects.
\end{abstract}

\begin{keyword}
Discrete Choice Experiment \sep Simulated Annealing \sep Multinomial logit model \sep Bayesian Optimal Design \sep MATLAB
\end{keyword}

\end{frontmatter}
\clearpage
\section{Required Metadata}

\begin{table}[h!]
\caption{Code metadata}
\centering

\small
\begin{tabular}{|l|p{6.2cm}|p{6.8cm}|}
\hline
\textbf{Nr.} & \textbf{Code metadata description} & \textbf{Please fill in this column} \\
\hline
C1 & Current code version & v1.0.0 \\
\hline
C2 & Permanent GitHub link to code/repository used for this code version
& \url{https://github.com/YichengMao98/DCEDesignSA} \\
\hline
C3 & Legal Code License & MIT License \\
\hline
C4 & Code versioning system used & Git (GitHub Releases) \\
\hline
C5 & Software code languages, tools, and services used & MATLAB R2025 \\
\hline
C6 & Compilation requirements, operating environments \& dependencies
& MATLAB-supported platforms (Windows, macOS, Linux) \\
\hline
C7 & If available Link to developer documentation/manual & \url{https://github.com/YichengMao98/DCEDesignSA} \\
\hline
C8 & Support email for questions & yicheng.mao1@ucalgary.ca\\
\hline
\end{tabular}

\label{tab:code-metadata}
\end{table}

\section{Motivation and significance}
\label{section1}
Discrete choice experiments (DCEs) are essential for analysing
stated preferences across various fields such as
marketing~\cite{liuInvestigatingRolePolitical2024}, health
care~\cite{clarkDiscreteChoiceExperiments2014,
lancsarConductingDiscreteChoice2008}, and
transportation~\cite{bliemerConstructionExperimentalDesigns2010}.
In each choice set, respondents are presented with a set of
alternatives defined by combinations of attribute levels associated
with the product or service being studied, and by observing their
selections, researchers can estimate preference parameters using
the Multinomial Logit (MNL) model~\cite{trainDiscreteChoiceMethods2003,
hensherCombiningSourcesPreference2000}.

As the design of experiments directly impacts quality of the statistical results, researchers often employ $\mathcal{D}$-optimal designs to ensure the accuracy of preference parameter estimation~\citep{huberImportanceUtilityBalance1996}.
These designs maximise the determinant of the information matrix under study, thereby improving the precision of predictions and the accuracy of parameter estimates under limited sample size conditions.
In the context of DCEs, the information matrix of the MNL model
depends on the values of the parameters to be estimated, which
are typically unknown at the experimental design stage. To
address this, Bayesian $\mathcal{D}$-optimal designs are adopted.
A Bayesian design specifies a prior distribution over the unknown
parameters and optimises the design criterion in expectation over
that distribution, thereby accounting for the uncertainty in the
parameter values~\citep{sandorDesigningConjointChoice2001}. Since the Bayesian
$\mathcal{D}$-optimality criterion admits no closed-form solution,
efficient search algorithms are required to construct these
designs, such as the modified Fedorov
algorithm (MF)~\citep{cookComparisonAlgorithmsConstructing1980}, coordinate exchange
(CE)~\citep{meyerCoordinateExchangeAlgorithmConstructing1995}, swapping algorithms (SWAP) \citep{huberImportanceUtilityBalance1996}, and simulated annealing
(SA)~\citep{maoConstructingBayesianOptimal2025}.

Although the CE algorithm has been widely adopted in the DCE
literature and demonstrated to yield superior computational
speed and statistical efficiency over the modified Fedorov
algorithm~\cite{kesselsEfficientAlgorithmConstructing2009}, it
remains a hill-climbing method that tends to converge prematurely
to local optima, particularly when the objective function is
complex and prior uncertainty is
large~\cite{maoConstructingBayesianOptimal2025}. DCEDesignSA addresses this by employing the SA algorithm,
which yields superior statistical efficiency compared to designs
generated via the CE algorithm in existing
packages, as demonstrated
through a systematic benchmark across multiple design scenarios,
including both full profile and partial profile
settings~\cite{maoConstructingBayesianOptimal2025,
maoSimulatedAnnealingModelRobust2026,MAO2025105305}.
\begin{table}[h]
\centering
\caption{Comparison of DCE design software tools and DCEDesignSA. The comparison is restricted to JMP's DCE-related functionality and, across all tools, to $\mathcal{D}$-optimal designs for the MNL model.}
\label{tab:comparison}
\footnotesize
\setlength{\tabcolsep}{4pt}
\renewcommand{\arraystretch}{1.15}
\resizebox{\textwidth}{!}{
\begin{threeparttable}
\begin{tabular}{lccccc}
\toprule
Feature
& JMP
& Ngene
& idefix
& choiceDes
& DCEDesignSA \\
\midrule
Software type
& Commercial
& Commercial
& R package
& R package
& MATLAB \\
GUI / interface
& Yes
& Yes
& No
& No
& Yes \\
Bayesian designs
& Yes
& Yes
& Yes
& No
& Yes \\
Partial profile designs
& Yes
& Yes
& No
& No
& Yes \\
Interaction effects
& Yes
& Yes
& No
& No
& Yes \\
Presentation order effects
& No
& No
& No
& No
& Yes \\
Optimisation algorithm
& CE
& SWAP / MF
& CE / MF
& MF
& SA \\
Survey implementation
& None
& None
& Shiny
& None
& Qualtrics \\
\bottomrule
\end{tabular}
\end{threeparttable}
}
\end{table}

Over the past few decades, several software tools have been developed to support the construction of DCE designs, differing in their optimisation algorithms, supported design structures, and implementation workflows, as summarised in Table~\ref{tab:comparison}. JMP~\cite{JMPDocumentation} and Ngene~\cite{ChoiceMetrics} provide graphical interfaces and support both Bayesian and partial profile designs. JMP supports utility specifications with second-order interaction effects and constructs designs using the CE algorithm, whereas Ngene offers a broader range of model specifications, coding options, and design constraints, and generates efficient designs using the SWAP and MF algorithms. Among R-based tools, idefix~\cite{traetsGeneratingOptimalDesigns2020} generates Bayesian $\mathcal{D}$-optimal designs for the MNL model and its variants using MF or CE algorithms, and includes a Shiny-based mechanism for presenting choice tasks and collecting empirical data. However, its core design-generation workflow remains code-based, and it does not support partial profile designs or interaction terms in the utility function. choiceDes~\cite{horneChoiceDesDesignFunctions2018} similarly provides code-based functions for generating local $\mathcal{D}$-optimal designs using an MF algorithm.

DCEDesignSA is designed to complement and extend this existing software ecosystem in three main respects. First, it integrates the DCE design workflow into an open-source graphical interface, reducing the need for users to specify designs through code and improving accessibility for researchers without programming expertise. Second, it implements SA-based optimisation for both full profile and partial profile $\mathcal{D}$-optimal designs, providing an alternative to the CE, SWAP, and MF algorithms used in existing tools. Third, it supports balanced profile order designs, allowing researchers to account for presentation order effects when the position of an alternative within a choice set may influence respondents’ choices~\cite{maoRandomizationDesignAnalysis2025}. In addition, DCEDesignSA exports completed designs directly to Qualtrics, automating the transition from design generation to survey implementation and avoiding the manual reformatting typically required when using other design-generation tools.

Taken together, these features make DCEDesignSA a practical platform for generating full profile, partial profile, and balanced profile order designs within a unified graphical workflow. Its main contribution is therefore not only the use of SA optimisation, but also the integration of design specification, optimisation, and Qualtrics-ready survey deployment in a single open-source tool.

\section{Software description}
\subsection{Software architecture}
\begin{figure}

    \centering
    \includegraphics[width=1\linewidth]{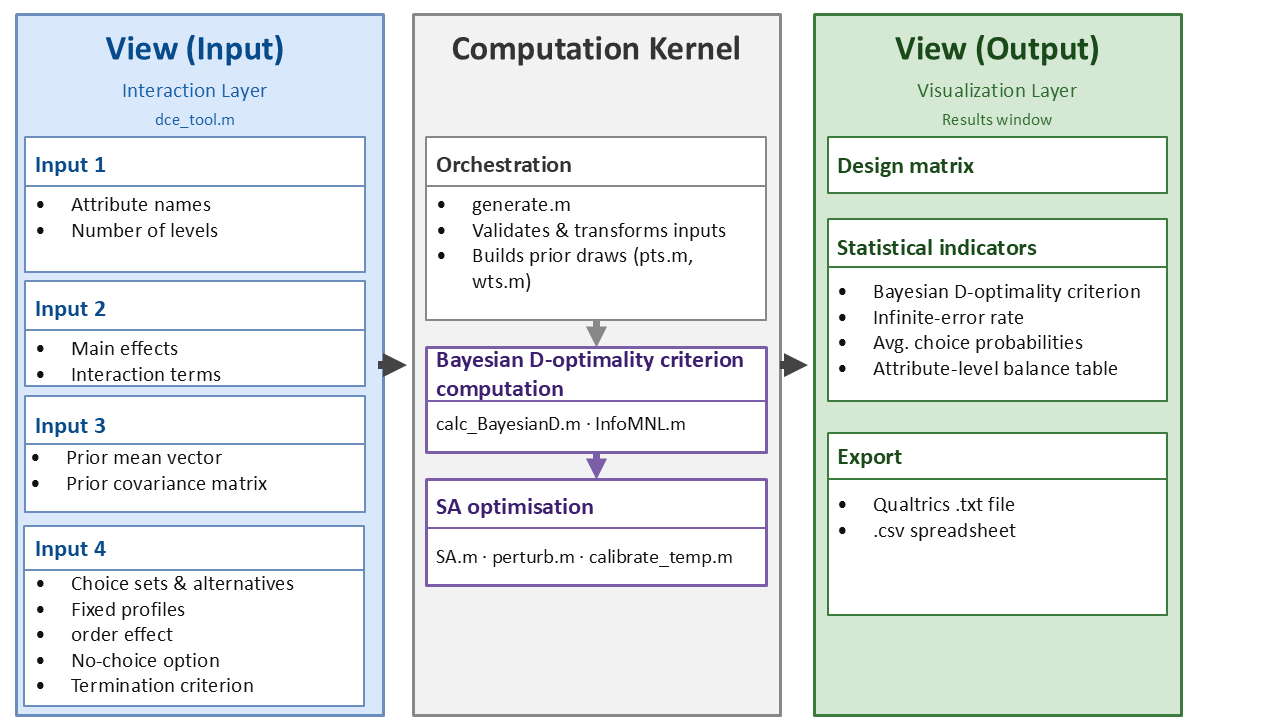}
    \caption{DCEDesignSA: Software Architecture}
    \label{architecture44}
\end{figure}

The developed software is implemented in MATLAB and features a GUI
built using MATLAB App Designer. As illustrated in
Fig.~\ref{architecture44}, DCEDesignSA is organised into four
functional layers:
\begin{enumerate}
    \item \textbf{User input}: this layer comprises four sequential
    steps, namely attribute definition, MNL model specification, prior
    elicitation, and design configuration.

    \item \textbf{Orchestration}: in this layer, \texttt{generate.m}
    accepts all user inputs and transforms them into a format suitable
    for Bayesian $\mathcal{D}$-optimality criterion calculation and
    optimisation.

    \item \textbf{Computation}: this layer consists of the Bayesian
    $\mathcal{D}$-optimality criterion computation module and the SA
    optimisation module.

    \item \textbf{Output}: this layer returns the Bayesian
    $\mathcal{D}$-optimal design matrix together with statistical
    indicators, including the Bayesian $\mathcal{D}$-optimality
    criterion, the infinite-error rate, and the average choice
    probabilities under the specified prior distribution. The
    infinite-error rate is defined as the proportion of prior draws
    yielding an infinite $\mathcal{D}$-error, where the
    $\mathcal{D}$-error is the inverse of the Bayesian
    $\mathcal{D}$-optimality criterion. The output layer also supports
    direct export to a Qualtrics-compatible \texttt{.txt} file and a
    \texttt{.csv} spreadsheet.
\end{enumerate}

\subsection{Software functionalities}
\subsubsection{Experiment Design Specification}

\begin{figure}[h!]
    \centering
    \includegraphics[width=1\linewidth]{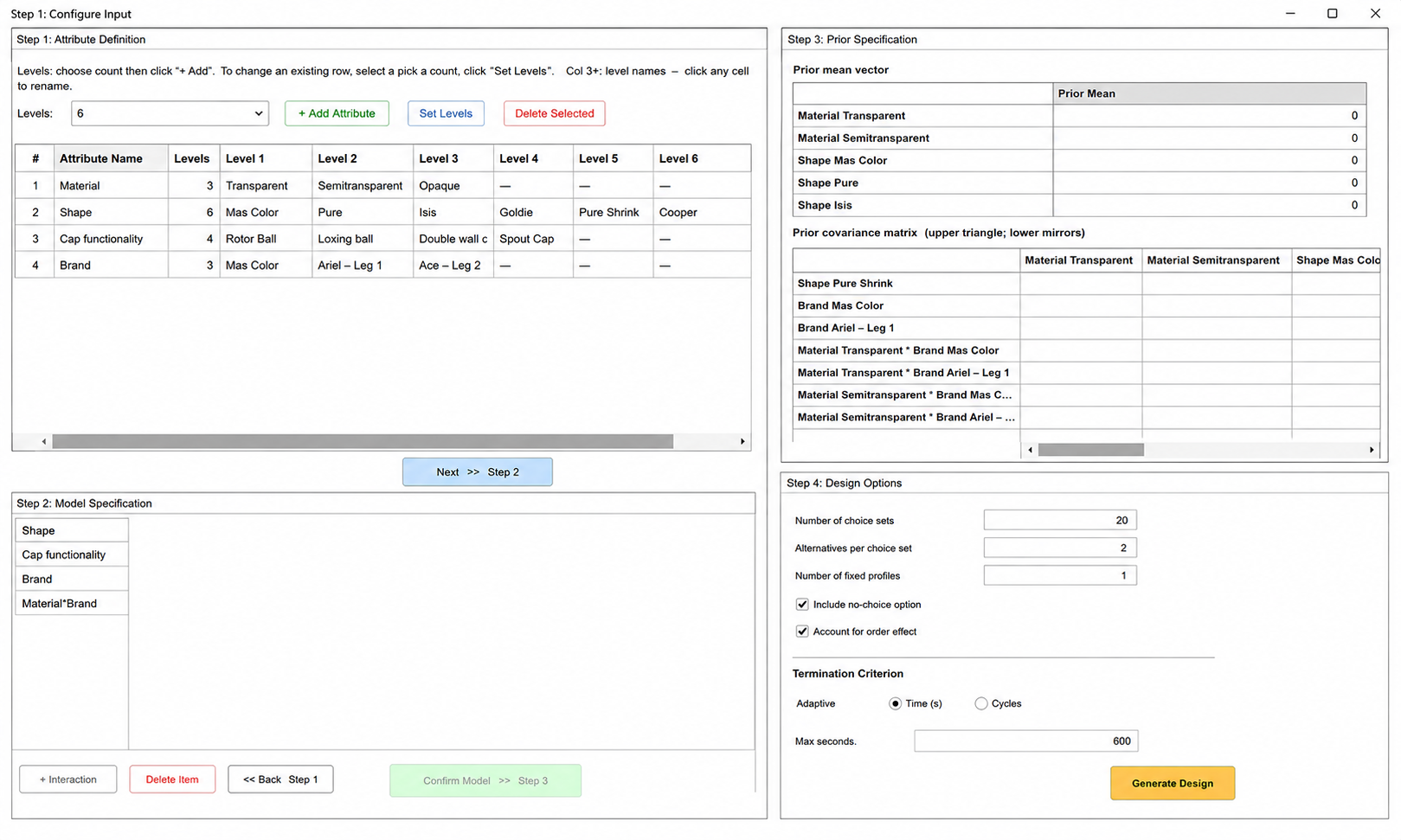}
    \caption{The DCEDesignSA graphical user interface showing
             the four-step configuration for the laundry detergent
             bottle design experiment.}
    \label{UI}
\end{figure}

As shown in Fig.~\ref{UI}, the GUI is organised into four input
panels:
\begin{enumerate}
    \item \textbf{Input 1: Attribute definition}. The user defines
    the attribute names, the number of levels for each attribute, and
    optional names for the attribute levels.

    \item \textbf{Input 2: Model specification}. The user selects the
    main effects and any interaction terms to be included in the
    choice model.

    \item \textbf{Input 3: Prior specification}. The user specifies
    the prior mean vector and covariance matrix used for the Bayesian
    $\mathcal{D}$-optimality criterion.

    \item \textbf{Input 4: Design configuration}. The user configures
    the number of choice sets, the number of alternatives per choice
    set, the number of fixed attributes for partial-profile designs,
    whether presentation-order effects are accounted for, and the SA
    termination criterion.
\end{enumerate}

\subsubsection{Statistical Objective Function and Optimisation Settings}
The software implements the MNL model as the underlying utility
framework~\cite{trainDiscreteChoiceMethods2003}. The utility that
a respondent attaches to profile $j$ $(j = 1, \ldots, J)$ in choice
set $s$ $(s = 1, \ldots, S)$ is defined as
\begin{equation}
\label{utility}
U_{js} = \mathbf{x}_{js}^{T}\boldsymbol{\beta} + \epsilon_{js},
\end{equation}
where $J$ and $S$ denote the numbers of profiles and choice sets
specified in Input~4, $\mathbf{x}_{js}$ is a $p \times 1$ vector
of attribute levels for profile $j$ in choice set $s$ as defined in
Input~1, and $\boldsymbol{\beta}$ is a $p \times 1$ vector of
preference parameters as specified in Input~2. Assuming the error
terms $\epsilon_{js}$ are independently and identically distributed
according to a Type-I extreme value distribution, the probability
that a respondent chooses profile $j$ in choice set $s$ is
\begin{equation}
P_{js} =
\frac{\exp \left( \mathbf{x}_{js}^{T}\boldsymbol{\beta} \right)}
{\sum_{i=1}^{J} \exp \left( \mathbf{x}_{is}^{T}\boldsymbol{\beta} \right)}.
\end{equation}
These choice probabilities enter the Fisher information matrix of
the MNL model, given by
\begin{equation}
\mathbf{M}(\mathbf{X},\boldsymbol{\beta})
=
\sum_{s=1}^{S}
\mathbf{X}_{s}^{T}
\left(
\mathbf{P}_{s}
-
\mathbf{p}_{s}\mathbf{p}_{s}^{T}
\right)
\mathbf{X}_{s},
\end{equation}
where $\mathbf{X} = (\mathbf{X}_{1}, \ldots, \mathbf{X}_{S})$ is
the model matrix over all choice sets, $\mathbf{p}_{s} = (P_{1s},
\ldots, P_{Js})^{T}$ is the vector of MNL choice probabilities for
choice set $s$, and $\mathbf{P}_{s} = \mathrm{diag}(\mathbf{p}_{s})$.
Since $\boldsymbol{\beta}$ is unknown at the design stage, a Bayesian
approach is adopted by integrating over the prior distribution
$\pi(\boldsymbol{\beta})$ specified by the user in Input~3, yielding
the Bayesian $\mathcal{D}$-optimality criterion
\begin{equation}
\label{DB}
\mathcal{D}_B =
\int_{\mathcal{B}}
\log\left|
\mathbf{M}(\mathbf{X},\boldsymbol{\beta})
\right|
\pi(\boldsymbol{\beta})
\, d\boldsymbol{\beta}.
\end{equation}
This integral is evaluated via the spherical-radial transformation
sampling method proposed by~\cite{gotwaltFastComputationDesigns2009},
which has been shown to outperform alternative numerical integration
approaches~\cite{yuComparingDifferentSampling2010}, with quadrature
points and weights provided as \texttt{.pts} and \texttt{.wts}
files within the package.  With $\mathcal{D}_B$ defined, Algorithm~\ref{alg:sa} describes the SA
procedure used to find the design that maximises $\mathcal{D}_B$,
where the candidate design is generated via the Exploration Rule
that adapts to different design contexts, including full profile,
partial profile, and balanced profile order design.

\begin{algorithm}[h!]
\caption{Simulated annealing procedure implemented in DCEDesignSA}
\begin{algorithmic}[1]
\Require Initial random design $\mathbf{X}$
\Ensure  Best design $\mathbf{X}_{Best}$ found by the algorithm

\State Set Initial Temperature $T_0$ using a random walk approach
\State Set iteration counter $k = 0$
\State Set $\mathbf{X}_{Best} = \mathbf{X}$
\State Record objective value $\mathcal{D}_B(\mathbf{X}_{Best})$

\While{Stopping Criterion not met}

    \State Update temperature: $T_k = \dfrac{T_0}{k+1}$
    \State Generate candidate design $\mathbf{X}'$ via the Exploration Rule
    \State Compute Bayesian $\mathcal{D}$-optimality criterion $\mathcal{D}_B(\mathbf{X}')$
    \State Compute acceptance probability:
    \[
        p = \min\left\{1,\; \exp\!\left(\frac{\mathcal{D}_B(\mathbf{X}') - \mathcal{D}_B(\mathbf{X})}{T_k}\right)\right\}
    \]

    \If{$\mathbf{X}'$ is accepted with probability $p$}
        \State Set $\mathbf{X} = \mathbf{X}'$
        \If{$\mathcal{D}_B(\mathbf{X}) > \mathcal{D}_B(\mathbf{X}_{Best})$}
            \State Update $\mathbf{X}_{Best} = \mathbf{X}$
        \EndIf
    \EndIf

    \State Increment iteration counter: $k = k + 1$

    \If{no new solutions are accepted in the last 1000 iterations}
        \State Reheat temperature: $T_k = T_0$
        \State Reset iteration counter: $k = 0$
    \EndIf

\EndWhile

\State \Return $\mathbf{X}_{Best}$

\end{algorithmic}
\label{alg:sa}
\end{algorithm}

As a practical guideline for termination criterion selection, the
\texttt{adaptive} criterion is recommended for simple design structures
with few attributes, up to three levels each, as it reliably converges within
minutes. For moderately complex designs, the \texttt{cycle} criterion
with fewer than 10 outer cycles provides a controlled runtime. For complex
designs, particularly those involving interaction terms or attributes
with more than four levels, the \texttt{time} criterion is recommended,
as the expanded design space may result in substantially longer runtimes
under adaptive termination.

\subsubsection{Result and Export}
Upon successful generation, a dedicated results window opens
automatically, organised into three tabs. The "Summary \& Balance"
tab displays the Bayesian $\mathcal{D}$-optimality criterion, total
runtime, the infinite-error rate, the design configuration flags,
and the frequency table reporting the count of each attribute level
across all regular alternatives. The "Design Matrix" tab presents
the optimised design. The "Choice Probabilities" tab reports the
average choice probability for each alternative in each choice set
under the specified prior distribution. At the bottom of the
results window, an export panel allows users to specify a filename
and export the design as a Qualtrics-compatible \texttt{.txt} file
or a \texttt{.csv} file.

\section{Illustrative examples}
\subsection{Input specification}
\begin{figure}
    \centering
    \includegraphics[width=1\linewidth]{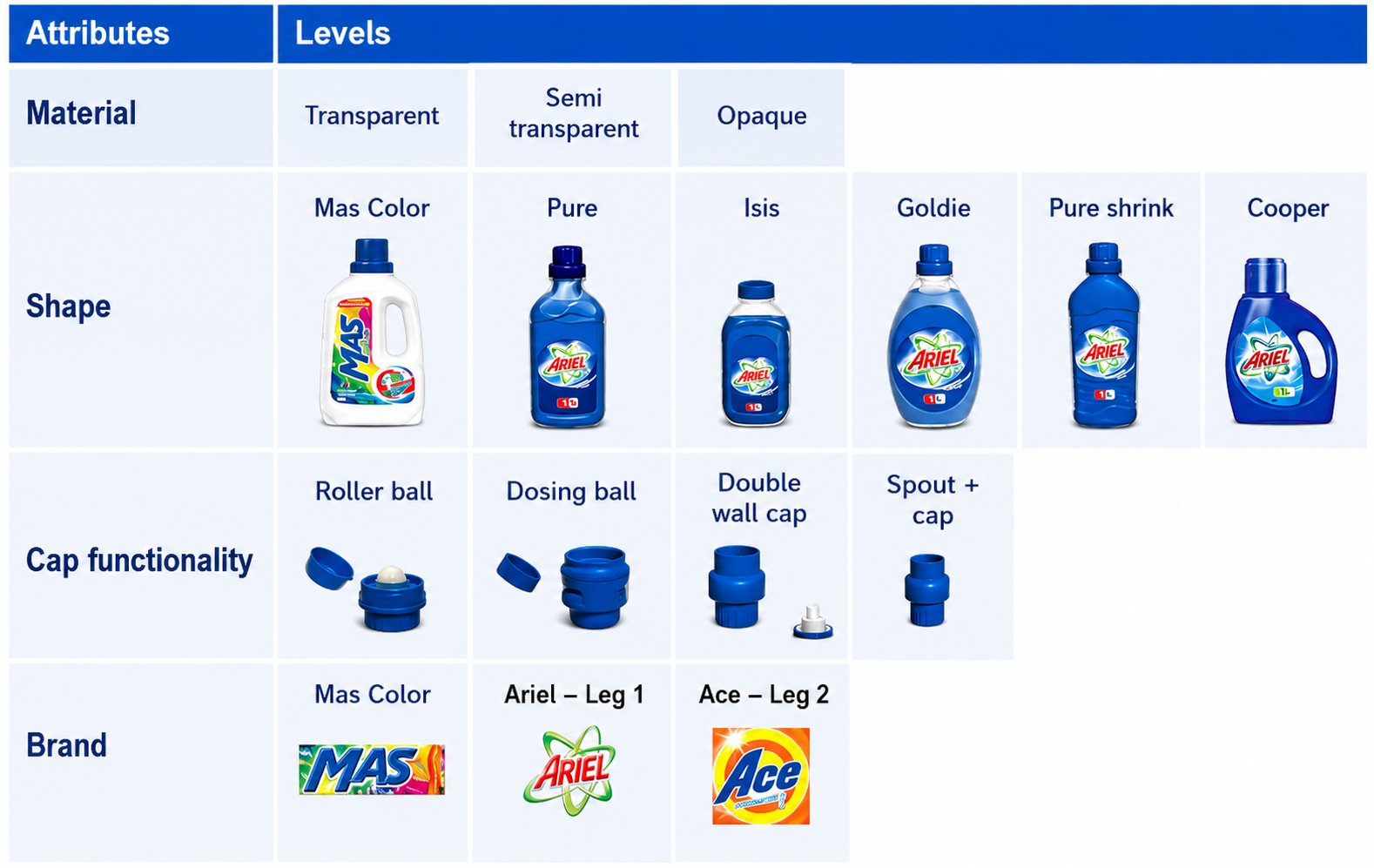}
    \caption{Laundry Detergent Bottle DCE Attributes Example}
    \label{ChoiceExample}
\end{figure}

To demonstrate the functionality of DCEDesignSA, we present an
application based on a real consumer study conducted by Procter~\&
Gamble (P\&G) to investigate preferences for laundry detergent
bottle designs among Mexican consumers. The experiment contained
four attributes, including shape, material, cap functionality, and
brand, with three to six levels each, as detailed in
Fig.~\ref{ChoiceExample}.

The design was configured via the GUI as shown in Fig.~\ref{UI}.
In Input~1, the four attributes and their respective levels were
defined as shown in Fig.~\ref{ChoiceExample}. In Input~2, a main
effect model was specified with one additional two-way interaction
term between Material and Brand. In Input~3, a zero prior mean
vector and an identity prior covariance matrix were specified for
all parameters, reflecting prior ignorance about consumer
preferences; an alternative-specific constant with a mean of $1$
was additionally specified for the opt-out alternative. In Input~4,
the design was configured with 26 choice sets, 2 alternatives per
choice set, and 1 fixed attribute, with both the no-choice option
and the presentation order effect activated. The termination
criterion was set to \texttt{time} with a limit of 600 seconds.

\subsection{Result and Qualtrics output}
\begin{figure}[h!]
    \centering
    \includegraphics[width=1\linewidth]{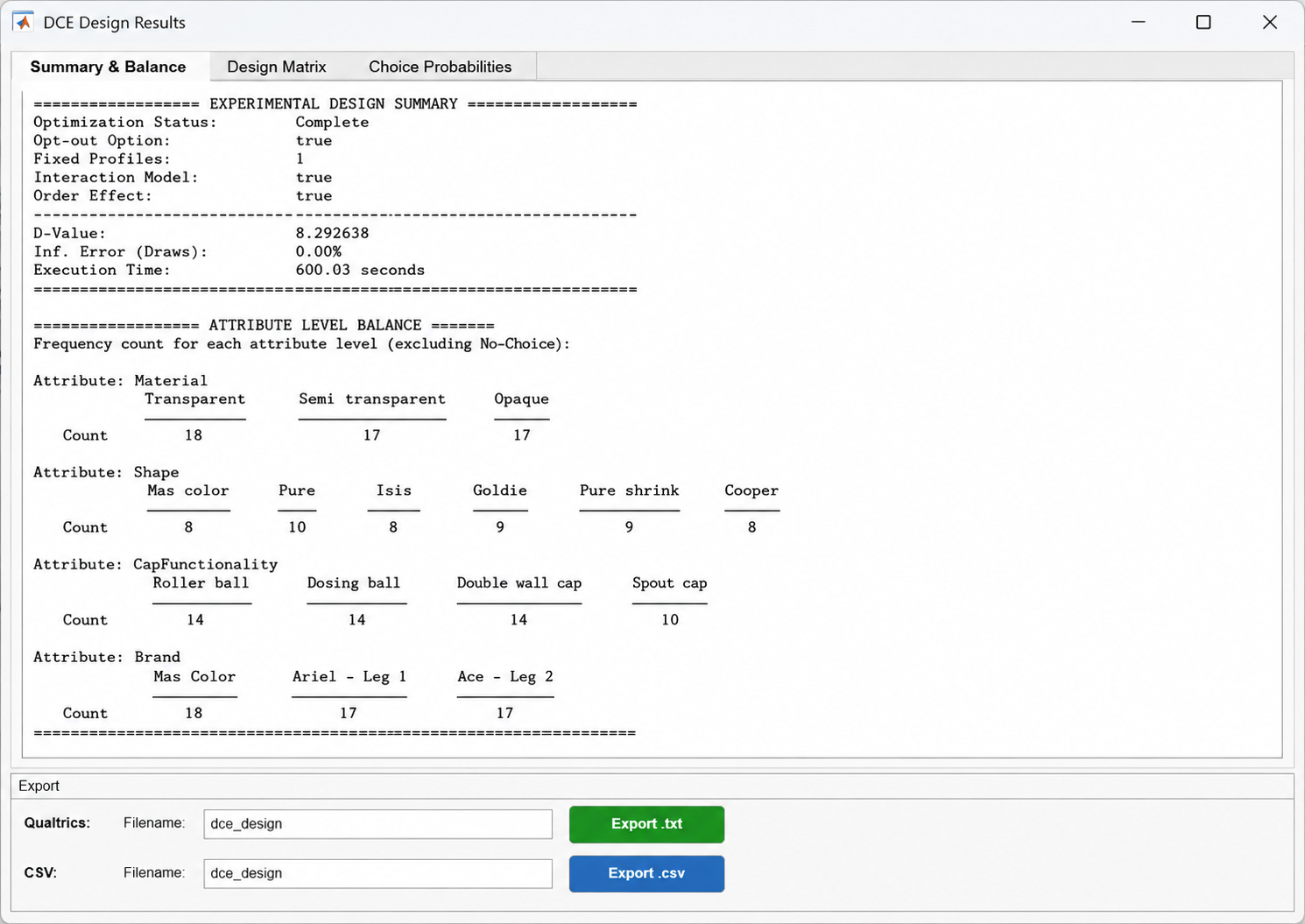}
    \caption{Summary and level balance output for the laundry
    detergent bottle design experiment (Bayesian
    $\mathcal{D}$-optimality criterion = 8.2926, infinite-error
    rate = 0.00\%).}
    \label{summary1}
\end{figure}

\begin{figure}[h!]
    \centering
    \includegraphics[width=1\linewidth]{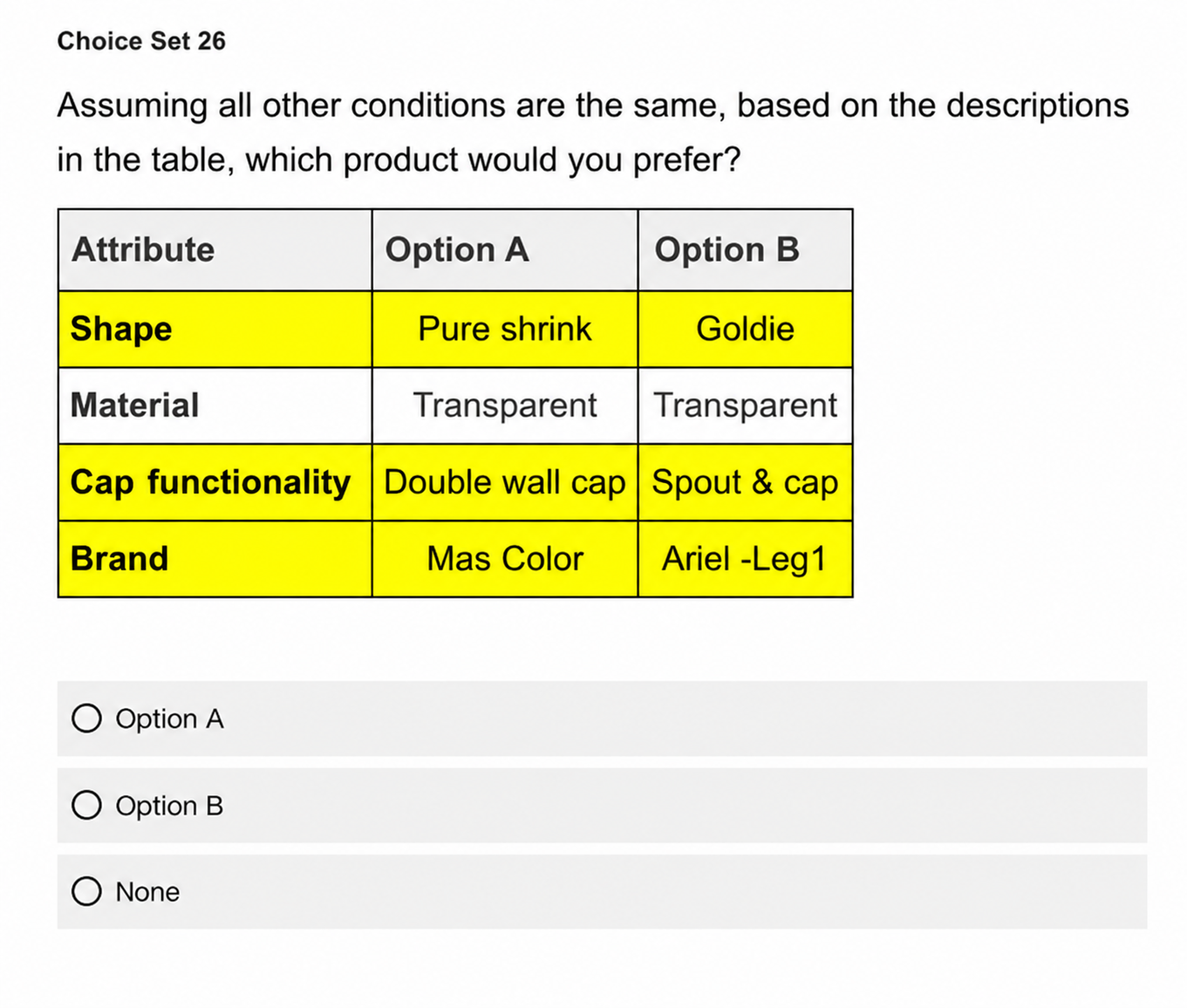}
    \caption{An example choice set displayed in the Qualtrics
    survey. Yellow highlighting indicates attributes that vary
    across alternatives.}
    \label{fig:qualtrics_choiceset}
\end{figure}

The resulting design achieved a Bayesian $\mathcal{D}$-optimality
criterion of $8.2926$, with an infinite-error rate of $0.00\%$,
indicating that the Fisher information matrix remained non-singular
across all prior draws throughout the optimisation, as summarised
in Fig.~\ref{summary1}. The design was subsequently exported to
Qualtrics for survey administration, and Fig.~\ref{fig:qualtrics_choiceset}
shows an example choice set as displayed to respondents.

\section{Impact}
As discussed in Section~2, existing DCE design tools offer complementary strengths but also leave a practical gap between methodological flexibility and ease of implementation. Commercial platforms such as JMP and Ngene provide GUI-based workflows, but they are closed-source and rely on proprietary software environments. R-based packages such as idefix and choiceDes offer open and flexible design-generation functions, but their main design workflows remain code-based. DCEDesignSA addresses this gap by combining an open-source MATLAB-based GUI with SA-based optimisation for Bayesian $\mathcal{D}$-optimal DCE design generation. The toolbox supports full profile and partial profile designs, interaction terms, opt-out alternatives, and balanced profile order designs, while also exporting designs directly to Qualtrics for survey deployment. In this way, DCEDesignSA reduces the technical burden of constructing statistically efficient DCE designs and streamlines the transition from design generation to empirical data collection.

\section{Conclusions}
DCEDesignSA is an open-source MATLAB toolbox that addresses a
critical gap in the DCE design literature by providing a
GUI-accessible, SA-based optimisation framework for generating
Bayesian $\mathcal{D}$-optimal designs. The toolbox supports full
profile and partial profile designs, interaction terms in utility,
order effects, and opt-out alternatives, with direct export to
Qualtrics-compatible \texttt{.txt} and \texttt{.csv} files for
immediate survey deployment.

\bibliographystyle{elsarticle-num}
\bibliography{references_clean_final}

\end{document}